%% file: qc_dim_sh_corr.tex
\documentstyle[prl,aps,multicol,epsf,epic,eepic]{revtex}

\begin{document}
\draft

\title{Topologically protected quantum bits from Josephson junction arrays}

\author{L.B.\ Ioffe$^{\ast,\ddagger}$, M.V.\
Feigel'man$^\ddagger$, A.\ Ioselevich$^\ddagger$, D.\
Ivanov$^\dagger$, M.\ Troyer$^\dagger$, and G.\ Blatter$^\dagger$}

\address{$^\ast$Department of Physics and Astronomy, Rutgers University,
Piscataway, NJ 08854, USA}
\address{$^\ddagger$Landau Institute for Theoretical Physics, 117940 Moscow,
Russia}
\address{$^\dagger$Theoretische Physik, ETH-H\"onggerberg, CH-8093 Z\"urich,
Switzerland}

\date{November 9, 2001}
\maketitle

\begin{multicols}{2}

\narrowtext

{\bf All physical implementations of quantum bits (qubits),
carrying the information and computation in a putative quantum
computer, have to meet the conflicting requirements of
environmental decoupling while remaining manipulable through
designed external signals. Proposals based on quantum optics
naturally emphasize the aspect of optimal isolation
\cite{Cirac-Zoller,Monroe,Turchette}, while those following the
solid state route exploit the variability and scalability of
modern nanoscale fabrication techniques
\cite{Loss,Shnirman,Averin,Mooij,Ioffe}. Recently, various designs
using superconducting structures have been successfully tested for
quantum coherent operation \cite{Nakamura_99,Friedmann_00,Wal_00},
however, the ultimate goal of reaching coherent evolution over
thousands of elementary operations remains a formidable task.
Protecting qubits from decoherence by exploiting topological
stability, a qualitatively new proposal due to Kitaev
\cite{Kitaev}, holds the promise for long decoherence times, but
its practical physical implementation has remained unclear so far.
Here, we show how strongly correlated systems developing an
isolated two-fold degenerate quantum dimer liquid groundstate can
be used in the construction of topologically stable qubits and
discuss their implementation using Josephson junction arrays.}




Any quantum computer has to incorporate some fault tolerance as we
cannot hope to eliminate all the various sources of decoherence.
Amazing progress has been made in the development of quantum error
correction schemes \cite{Preskill} which are based on redundant 
multi-qubit encoding of the quantum data combined with error 
detection- and recovery steps through appropriate manipulation 
of the data.  Error correction schemes are generic (and hence 
are applicable to any hardware implementation),
but require repeated active interference with the computer during
run-time; the delocalization of the data, often in a hierarchical
structure, boosts the system size by a factor $10^2$ to $10^3$.
Delocalization of the quantum information is also at the heart of
topological quantum computing \cite{Kitaev}, however, the
stabilization against decoherence is entirely deferred to the
hardware level (hence it is tied to the specific implementation)
and is achieved passively. In searching for a physical
implementation of topological qubits  one strives for an extended
(many body) quantum system where the Hilbert space of quantum
states decomposes into mutually orthogonal sectors, each sector
remaining isolated under the action of local perturbations.
Choosing the two qubit states from groundstates in different
sectors protects these states from unwanted mixing through noise;
protection from leakage within the sector has to be secured
through a gapped excitation spectrum. As no local operator can
interfere with these states, global operators must be found (and
implemented) allowing for the manipulation of the qubit state.

A promising candidate fulfilling the above requirements is the
quantum dimer system \cite{Kivelson_87,Rokhsar_88,Wen_91}: recent
quantum Monte Carlo simulations of the dimer model on a triangular
lattice provide evidence for a gapped liquid groundstate
\cite{Moessner_00} (see \cite{Misguich_99,Sachdev_92} for a
discussion of similar exotic groundstates in spin models) and we
will discuss its topological robustness below. The physical
implementation of such a dimer system can be realized with the
help of quantum Josephson junction arrays. Such arrays allow for a
proper `material design' as the energy scales governing the charge
and phase degrees of freedom can be tuned through junctions with
appropriate tunnel barriers and capacitances. Shaped into the form
of a ring, the Josephson junction array emulating a dimer system
allows for the construction of topologically protected qubits.

\begin{figure}
  \centerline{\epsfxsize = 7.6cm \epsfbox{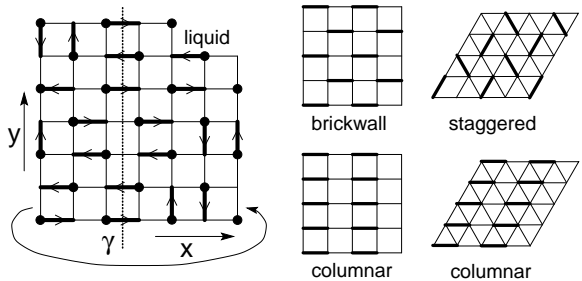}}
  \narrowtext\hspace{0.0cm}
  \caption{Dimers on a square lattice forming liquid (main panel),
  brickwall, and columnar configurations. Periodic boundary
  conditions are applied along the $x$-direction. The sublattice $A$ is
  marked by solid bullets; dimers are attributed a direction pointing
  away from bullets. The topological invariant $\phi$ counts the
  number of dimers crossing the reference line $\gamma$, with
  dimers pointing to the right (left) counting with positive
  (negative) signs (here, $\phi = 1$). Right: staggered and
  columnar configurations in the triangular lattice.}
  \label{fig:dimers}
\end{figure}

The simplest dimer model is defined on a square lattice; allowed
configurations are coverings with dimers connecting neighboring
vertices satisfying the constraint that every vertex belongs to
only one dimer, see Fig.\ \ref{fig:dimers}. Attributing
equal energies to these states, the classical model is
characterized by an extensive entropy. The quantum dynamics is
introduced by the hopping Hamiltonian
\begin{equation}
   H_t = -t \sum_{\Box}
   \bigl(|\input{vv.eepic}\rangle
   \langle \input{hh.eepic}|
   +|\input{hh.eepic}\rangle
   \langle \input{vv.eepic}|\bigr)
   \label{H_t}
\end{equation}
rotating parallel dimers on appropriate plaquettes $\Box$. As a
result, the degeneracy of the classical model is lifted, with
`columnar' and `staggered' (e.g., brickwall) phases (see Fig.\
\ref{fig:dimers}) assuming minimal and maximal energies,
respectively. In order to avoid such crystalline ordering we
frustrate parallel dimers with the interaction term
\begin{equation}
   H_v = v \sum_{\Box}
   \bigl(|\input{vv.eepic}\rangle \langle\input{vv.eepic}|
   +|\input{hh.eepic}\rangle \langle\input{hh.eepic}| \bigr).
   \label{H_v}
\end{equation}
At $t=v$ (the Rokhsar-Kivelson point) a zero energy dimer
liquid groundstate is formed involving all allowed
configurations with equal amplitudes \cite{Rokhsar_88}.
For the square lattice this liquid phase is restricted
to the point $t=v$ and tuning $v$ away from $t$ the liquid
loses the competition with various crystalline phases
\cite{Fradkin}. Here, we are interested in systems developing a
stable liquid over an extended parameter range. Such a
stabilization may be achieved by the term
\begin{equation}
   H_{d} = \sum_{\Box}
   \bigl[-t\bigl(\,|\input{hh2.eepic}\rangle\langle\input{dd1.eepic}|
   +|\input{vv2.eepic}\rangle \langle\input{dd2.eepic}|
   +h.c. \bigr)
   +\mu |\input{d.eepic}\rangle \langle\input{d.eepic}|\,\bigr]
   \label{H_t'}
\end{equation}
transforming horizontally ($|\input{hh2.eepic}\rangle$) and
vertically ($|\input{vv2.eepic}\rangle$) shifted dimers into
diagonal ones $|\input{dd2.eepic}\rangle$ at the energy cost
$2\mu$ (only one family of diagonals is chosen). Tuning the
chemical potential $\mu$ from $\infty$ to 0, diagonal dimer flips
proliferate and we can study the evolution of the dimer model as
we go from a square to a triangular lattice \cite{MSF}.
\vspace{-0.3cm}
The above type of quantum dimer models has been actively discussed
in the context of short-range resonating-valence-bond (RVB) models
\cite{Anderson} for high temperature superconductivity
\cite{Kivelson_87}. In order to gain more insight into its
physical properties we exploit a mapping to an `electrodynamic'
analogue \cite{Fradkin,MSF,Ioffe_89}: We decorate the bipartite
square lattice (sublattices A and B) with static charges
$\rho_\Box = \pm 1$, placing all positive charges on the
sublattice A, see Fig.\ \ref{fig:dimers}. A dimer covering $\{{\bf
d}_{\langle i,j\rangle}\}$ of bonds $\langle i,j \rangle$ of this
`ionic crystal' is mapped to a distribution of electric dipoles
$\{-{\bf p}_{\langle i,j\rangle}\}$; the electric field ${\bf e} =
- \langle {\bf p}_{\langle i,j\rangle} \rangle_{\delta A}$
resulting from averaging over small areas $\delta A$ then
satisfies Gauss' law $\nabla {\bf e} = \langle \rho\rangle_{\delta
A} = 0$. The simplest dynamics preserving this constraint is given
by Faradays law or, more precisely, (compact) quantum
electrodynamics. Here, we are interested in the large-scale
(global) aspects of the model and hence we will keep the
discussion on a phenomenological level; the physical properties of
the dimer liquid phase will be described in terms of the electric
field ${\bf e}$ and the dielectric constant~$\varepsilon$.

While the order characterizing the crystal phases is obvious, the
topological order present in the dimer liquid phase is more subtle
\cite{Kivelson_87,Rokhsar_88,Wen_91}: we consider a cylindrical
geometry with periodic boundary conditions along the $x$-axis,
leaving the boundaries along the $y$-direction free (an additional
periodic boundary condition along the $y$-direction defines a
torus, see later). In the absence of diagonal dimers (the limit
$\mu \rightarrow \infty$ in (\ref{H_t'})) the sum $\phi$ counting
directed dimers cut by the reference line $\gamma$ parallel to the
$y$-axis defines a topological order parameter (dimers pointing
into opposite directions are counted with opposite sign): the
action of the Hamiltonian $H=H_t+H_v$ leaves $\phi$ invariant and
the Hilbert space splits into topological sectors ${\cal H}_\phi$
characterized by the integers $\phi$. Within our electrodynamic
analogue this topological invariant is given by the flux $\phi =
\int dy\, \varepsilon e_x$ of the electric induction across
$\gamma$ and each sector ${\cal H}_\phi$ is characterized by an
average polarization $e_x = \phi/\varepsilon L_y$ producing a
groundstate energy $E = \varepsilon e^2 L_x L_y/2 = \phi^2
L_x/2\varepsilon L_y\equiv E_\phi$. At finite values of the
chemical potential $\mu$ the Hamiltonian (\ref{H_t'}) introduces
mobile defects with charge $\pm 2$ \cite{MSF,FradkinShenker}.
These charges mix states with electric flux $\phi \pm 2$: creating
two defects of charge $\pm 2$, carrying one defect around the
cylinder, and subsequent annihilation with its partner changes the
flux $\phi$ by 2; as a result, the infinite set of sectors
${\cal H}_\phi$ collapses to the even and odd sectors ${\cal H}_{e,o}$.

The Hamiltonian projected onto the space of ground states within
topological sectors ${\cal H}_\phi$ takes the form $H_{\rm sec}
= \sum_\phi \bigl(E_\phi |\phi \rangle \langle \phi|
+ J (|\phi\rangle \langle \phi+2| + h.c.) \bigr)$. For large
values of the chemical potential $\mu \gg t$ the defect
pairs are only virtual and the amplitude $J$ mixing states with
different $\phi$ is exponentially small in the circumference
$L_x$, $J \propto (t/\mu)^{L_x}$.

As $\mu$ drops below an energy of order $t$, defects proliferate
and undergo Bose condensation; the concomitant long range order in
the condensate phase $\varphi$ implies large charge fluctuations
and thus the electric field flux across $\gamma$ fluctuates
strongly. In this phase the mixing between sectors is large with
$\langle \phi^2\rangle \propto L_y$: assuming a superfluid density
$\rho_s$, the hydrodynamic description of the charged superfluid
\cite{Ioffe_89} is given by the Euclidean action (in Fourier
space; $\rho$ denotes the conjugate to $\varphi$) $S_{\rm
\scriptscriptstyle E} = \sum_{\omega,{\bf K}} \bigl[\omega\,
\rho_{-\omega,-{\bf K}} \varphi_{\omega,{\bf K}}+\rho_s
K^2|\varphi_{\omega,{\bf K}}|^2/2 +|\rho_{\omega,{\bf K}}|^2
/2\varepsilon K^2\bigr]$, from which we derive a density
correlator $\langle \rho\rho\rangle = \rho_s K^2/(\rho_s/
\varepsilon+\omega^2)$. The relation $\varepsilon\nabla\cdot{\bf
e}=\rho$ allows us to find the equal time correlator of the
electric induction $\langle \varepsilon e_\mu({\bf R},t)
\varepsilon e_\nu({\bf R'},t)\rangle =\delta_{\mu\nu}\delta^2({\bf
R}-{\bf R'}) \sqrt{\rho_s \varepsilon}/2$, resulting in the mean
electric flux $\langle \phi^2 \rangle = \sqrt{\rho_s\varepsilon}
L_y/2$ crossing the line $\gamma$; the local correlations in the
electric field are a consequence of the dispersion-free plasmon
spectrum.

The new groundstates $|e,o\rangle \propto \sum_{\phi_{e,o}}
a_{\phi_{e,o}} |\phi_{e,o}\rangle$ involve contributions from
all sectors ${\cal H}_{e,o}$, with $\phi_e=2k$ and
$\phi_o=2k+1$ running over even and odd integers; their
physical properties are conveniently characterized by the
distribution function $P(\phi)=|a_\phi|^2 \propto
\exp(-\phi^2/ 2\sigma)$, with the Gaussian form
appropriate for a process involving a sum of statistically
independent elements. The width $\sigma$ is obtained through a
comparison with the second moment $\langle \phi^2\rangle$, $\sigma
= \sqrt{\rho_s\varepsilon}L_y/2$. Expectation values in even
and odd sectors ${\cal H}_{e,o}$ are given by the sums over
even and odd $\phi$ with weight $P(\phi)$. As a result,
differences between expectation values in ${\cal H}_{e,o}$
are exponentially small in $L_y$, e.g., the average squares
of electric fluxes differ by
\begin{equation}
   \langle \phi^2 \rangle_e - \langle \phi^2 \rangle_o \sim
   \exp(-\pi^2\sqrt{\rho_s \varepsilon}L_y/4).
\label{phi_eo}
\end{equation}
Hence, the presence of diagonal dimers provides us with only two
topological sectors for a system defined on a cylinder
(and 4 sectors on a torus). The corresponding groundstates are
degenerate in the limit $L_{x,y}\to \infty$. Finite-size effects
as well as local disorder lead to an exponentially weak energy
splitting (cf.\ (\ref{phi_eo})). Furthermore, the topological
property of the system inhibits any mixing of the two groundstates.
In the thermodynamic limit
the collapse of infinitely many to two topological sectors
involves a quantum phase transition in the parameter $\mu$.
For $\mu=0$ the above model corresponds to a symmetric triangular
lattice (upon including the $v$-term) and we expect
favorable conditions for the construction of protected qubits.

Indeed, this expectation is supported by results obtained from
numerical studies: recent Monte Carlo simulations by Moessner and
Sondhi \cite{Moessner_00} on large systems with $L_{x,y}=36$
exhibit short-range dimer-dimer correlations within a parameter
region $2/3 < v/t < 1$, indicative of a liquid state. Furthermore,
the weak temperature dependence of this data within the
temperature interval $0.25 t > T > 0.03 t$ provides evidence for a
gapped spectrum (increasing $v$ beyond $t$, a first-order
transition takes the liquid into the staggered phase, while 
decreasing $v$ below $2t/3$ establishes some crystalline order).

We have numerically diagonalized the $t$-$v$ dimer Hamiltonian $H
= H_t+H_v$ on a triangular lattice in order to estimate the gap
protecting the liquid and verify the absence of low lying edge
states; in addition we have investigated the robustness of the
degenerate groundstates under local perturbations (requiring exact
diagonalization limited to smaller systems). We have chosen both
cylindrical and toroidal geometries, with system sizes going up to
$L_{x,y} = 6$; although subject to finite size effects we expect
that our approach describes well the liquid phase where the
short-ranged dimer correlations (of order of one
lattice constant \cite{Moessner_00}) reduce the influence of the
boundaries. Our exact diagonalization study confirms the presence
of a gap of order $\Delta \approx 0.1 t$, taking up large values
on approaching the Rokhsar-Kivelson point $t=v$. Comparing the
spectra for tori (no open boundaries) with those of cylinders
(where the two open boundaries could accommodate edge states,
would they exist) we note that the gap persists, from which
we infer the absence of low lying edge-states in the dimer
liquid (this should be contrasted with the situation in the
quantum Hall system).

\begin{figure}
  \centerline{\epsfxsize = 6.0cm \epsfbox{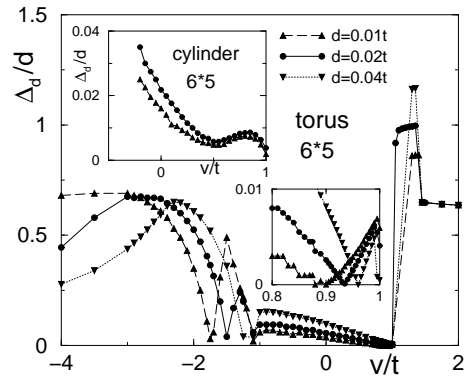}}
  \narrowtext\vspace{-0.1cm}
  \caption{Splitting $\Delta_d$ of the groundstate energies
  under the action of a disorder potential of strength $d$
  for a $6*5$ torus and a $6*5$ cylinder (inset; periodicity is
  along $x$ with $L_x = 6$).
  The disorder only weakly perturbs the dimer liquid groundstates
  at $v < t$ (note the expanded view near $v\sim t$ for the
  $6*5$ torus). The crossovers between various phases
  become clearly visible in the susceptibility to a disorder
  potential.}
  \label{fig:dis_tc}
\end{figure}

In order to test the susceptibility of the degenerate groundstates
to local perturbations we assign all links random chemical
potentials $\mu_d$ homogeneously distributed over the interval
$[-d/2,d/2]$, $d< t$. In finite cylinders and tori, particular 
symmetries specific to some geometries $L_{x,y}$ produce an exact
degeneracy of eigenstates from different topological classes;
for the $6*5$ cylinder and torus used
in our study the symmetry analysis predicts double-degeneracy.
As shown in Fig.\ \ref{fig:dis_tc} the disorder induced
groundstate splitting $\Delta_d$ collapses dramatically as we go
from the crystal ($v>t$) to the liquid phase ($v < t$) and slowly
recovers as $v$ is decreased further. We attribute the sharp drop
at $v\approx t$ to the first-order solid-liquid transition; the
small disorder splitting at $v<t$ testifies for the efficient
protection of the dimer liquid groundstates from local
perturbations. Decreasing $v$ further below $t$ we observe level
crossings which we attribute to the appearance of intermediate
crystalline phases \cite{Moessner_00} before reaching the
fluctuating columnar phase at large negative values of $v$.
\vspace{-0.3cm}
Summarizing, the combination of Monte Carlo simulations
\cite{Moessner_00} and diagonalization indicate that on
approaching the point $v=t$ from below, the triangular dimer model
develops an isolated degenerate dimer liquid groundstate free of
edge states. Within the parameter region $\sim 0.8 < v/t <1$ we
find a gap of order $0.1 t$ and groundstates which are robust
under disorder to within 1\% of the disorder potential; we expect
a further increase of this robustness with system size, cf.\
(\ref{phi_eo}). Translating these findings to potential
implementations of this two-state system, we find the latter to be
quite tolerant with respect to (static) local variations in
parameters appearing in the fabrication process. Regarding
decoherence, the amplitude mixing of the two groundstates is
strongly suppressed, since it involves creation of topological
defects (real or virtual) violating the dimer constraint. The
decoherence in the relative phase of the two groundstates
originates either from their adiabatic splitting by an external
low-frequency noise, or the creation of non-topological
excitations within each sector. The former is suppressed for the
same reason that leads to the robustness with respect to local
disorder, while the latter is inhibited by the gap $\Delta$, which
sets the requirement $T\ll\Delta$ on the operation temperature.

\begin{figure}
  \centerline{\epsfxsize = 8.0cm \epsfbox{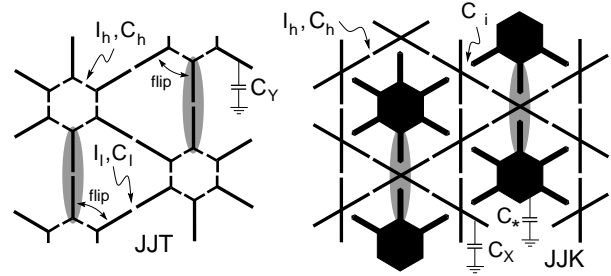}}
  \narrowtext\hspace{0.0cm}
  \caption{Josephson junctions arrays forming triangular (JJT)
  and Kagome (JJK) lattices and emulating dimer systems. JJT:
  The large capacitance $C_h$ joins hexagonal islands into
  electric units which are frustrated by a global gate to
  accept 1/2 Cooper pair. Cooper pairs resonating between
  hexagonal islands form dimers (shaded ellipses); dimer
  flips due to $H_t$ are indicated by arrows. JJK:
  Cooper pairs on $X$-shaped islands are capacitively coupled
  through star-shaped islands. The gate is tuned to allow
  for 1/2 Cooper pair per hexagon. Dimers are pairs residing
  on $X$-shaped islands and polarizing adjacent star-shaped
  islands.}
  \label{fig:JJA}
\end{figure}

In the following we propose two types of Josephson junction arrays
for the implementation of the above topologically protected
quantum dimer liquid groundstates and their use for quantum
computing. In the first array (JJT), the vertices of the
triangular lattice are structured into six Y-shaped
superconducting islands with two ends joining into the hexagonal
vertex and the third end linking to the neighboring hexagon, see
Fig.\ \ref{fig:JJA}. The 5 parameters, the capacitance
$C_{\rm\scriptscriptstyle Y}$ of the island to the groundplate,
the capacitance $C_h$ and Josephson current $I_h$ of the hexagonal
vertex junction, and the capacitance $C_l$ and Josephson current
$I_l$ of the link junction, are chosen to emulate the triangular
quantum dimer model. All capacitances and currents define
corresponding energies $E^{\rm\scriptscriptstyle C}= e^2/2C$ and
$E^{\rm\scriptscriptstyle J}= \Phi_0 I/2\pi c$, where $\Phi_0 =
hc/2e$ denotes the flux quantum (e.g., $E_h^{\rm\scriptscriptstyle
C}=(2e)^2/2C_h$, $E_h^{\rm \scriptscriptstyle J}=\Phi_0 I_h/2\pi
c$). We first find $C_{\rm \scriptscriptstyle Y}$, $C_h$, and
$I_l$ defining the classical dimer states: We choose a large
capacitance $C_h$ in order to join the islands electrically into
one hexagonal vertex. A small capacitance $C_{\rm
\scriptscriptstyle Y}$ defines the large charging energy $E_{\rm
hex} \approx E^{\rm \scriptscriptstyle C}_{\rm \scriptscriptstyle
Y}/6$ of the hexagonal vertex, the basic energy scale of the
array. We introduce `charge frustration' of the vertices by
biasing the array with a global electric gate such as to equalize
the energies (to an accuracy better than $E^{\rm\scriptscriptstyle
J}_l$) of two states differing by one Cooper pair (this defines a
`half-filled' array); the large charging energy $E_{\rm hex}$
lifts other charge states to high energies. The Cooper pairs lower
their energy via tunneling (involving the coupling $I_l$) through
the link junction joining two vertices --- the `bonding' state of
such a Cooper pair defines the dimer state. With only half a
Cooper-pair available per hexagonal vertex, each vertex is
involved in the formation of one and only one of these `valence
bonds'. This defines the `classical' configurations of the dimer
model; the corresponding Hilbert space of dimer states is
protected by the energy scale $E^{\rm\scriptscriptstyle J}_l$.

The dimer dynamics involves the vertex junction and proceeds via
localization of one dimer to the vertex (with an energy cost
$E^{\rm\scriptscriptstyle J}_l$) and subsequent hops (of two
parallel dimers/Cooper pairs over junctions with energy
${E^{\rm\scriptscriptstyle J}_h}$, see Fig.\ \ref{fig:JJA}) to the
new vertex islands. The hopping amplitude is of order $t \sim
{E^{\rm\scriptscriptstyle J}_h}^2/E^{\rm \scriptscriptstyle J}_l$;
explicit calculation gives the result $t=(9/16){E^{\rm
\scriptscriptstyle J}_h}^2/E^{\rm \scriptscriptstyle J}_l$. The
electrostatic interaction between dimers depends on the choice of
the capacitance matrix and its dependence on the dimer
configuration is non-trivial. First, we compare the energies of
the staggered and columnar configurations and find the interaction
energy $v$ between parallel dimers; in the limit $C_l \ll
C_{\rm\scriptscriptstyle Y},C_h$, $v = E^{\rm\scriptscriptstyle
C}_h(C_l/C_h) [(1+C_{\rm\scriptscriptstyle Y}/C_h)
(1+3C_{\rm\scriptscriptstyle Y}/C_h)]^{-2}$. Second, we have
checked that the electrostatic energies of the liquid phase dimer
configurations indeed scale (to within $\sim 5 \%$ accuracy) with
the number of parallel dimer pairs. The condition $C_l \ll
C_{\rm\scriptscriptstyle Y}$ guarantees a short range interaction
between dimers.
\vspace{-0.3cm}

The backbone of the second array (JJK) is made from X-shaped
islands arranged into a Kagome lattice with capacitive and
Josephson couplings $E^{\rm\scriptscriptstyle C}_h$ and
$E^{\rm\scriptscriptstyle J}_h$, see Fig.\ \ref{fig:JJA}. A second
triangular lattice of star-shaped islands introduced into the
hexagons of the Kagome lattice is only capacitively coupled (with
energy $E^{\rm\scriptscriptstyle C}_i$) to the islands of the
Kagome lattice; $C_{\rm\scriptscriptstyle X}$ and $C_{\ast}$
denote the capacitances of the islands to the ground. The Kagome
sites are biased with a global external gate to accept one
additional Cooper pair per hexagon at equal energy, resulting in a
filling with 1/2 Cooper pair per hexagon. The similarity between
the Kagome lattice and the triangular array becomes obvious when
joining pairs of Y-shaped islands (corresponding to the limit
$C_l\rightarrow\infty$ and $I_l\rightarrow\infty$) and contracting
the link to obtain the X-shaped island --- the previous vertex
junctions now play the role of the junctions on the Kagome
lattice, hence the same index `$h$' has been chosen for the
junctions on the hexagons; note that the present array involves
only one type of Josephson junctions. By analogy, dimers now
correspond to Cooper pairs localized onto the X-islands of the
Kagome lattice. As dimers should not touch one another, no two
bosons are allowed on the same hexagon: We choose small
capacitances $C_h$ in order to isolate hexagons from one another
(note the difference to the JJT array) and large capacitances
$C_i$ to join the six X-shaped islands forming the hexagon into
one electrostatic unit via their strong coupling to the central
star-shaped island. The capacitances $C_{\rm\scriptscriptstyle
X}$, $C_{\ast}$, and $C_i$ then define the charging energy
required to put two Cooper pairs on the same hexagon, $E_{\rm hex}
\approx (C_i/C_{\rm \scriptscriptstyle X})^2
E_\ast^{\rm\scriptscriptstyle C}$, which is our basic energy scale
(we assume $C_i < C_\ast, C_{\rm\scriptscriptstyle X}$ in order to
guarantee good screening on large distances).

The motion of the Cooper pairs (the dimer dynamics) involves the
hopping amplitude $E^{\rm \scriptscriptstyle J}_h$ and the
charging energy $E_{\rm hex}$ of the virtual state with two Cooper
pairs on the same hexagon, $t \sim {E^{\rm \scriptscriptstyle
J}_h}^2/E_{\rm hex}$; note that we require $E^{\rm
\scriptscriptstyle J}_h \ll E_{\rm hex}$ not to perturb the proper
frustration of the array. The interaction energy $v$ again derives
from a comparison between the staggered and columnar states and
has to meet the conflicting requirements of increasing the energy
of parallel dimers (next-nearest neighbor interaction) while
leaving non-parallel dimer pairs (next-next-nearest neighbor
interaction) approximately unperturbed. In order to produce a
positive $v$ we need a finite coupling $C_h$, which in turn
reduces the basic energy $E_{\rm hex}$. We find that the set $5
C_i= C_\ast = C_{\rm\scriptscriptstyle X}=20 C_h$ produces a
(renormalized) protective energy $E_{\rm hex}^r \approx 0.008
E_\ast^{\rm\scriptscriptstyle C} \approx 0.2 E_{\rm hex}$ and a
repulsive energy $v \approx 0.1 E_{\rm hex}^r$. Choosing $E^{\rm
\scriptscriptstyle J}_h \approx 0.3 E_{\rm hex}^r$ we can properly
satisfy the condition $t \approx v$.

\begin{figure}
  \centerline{\epsfxsize = 7.8cm \epsfbox{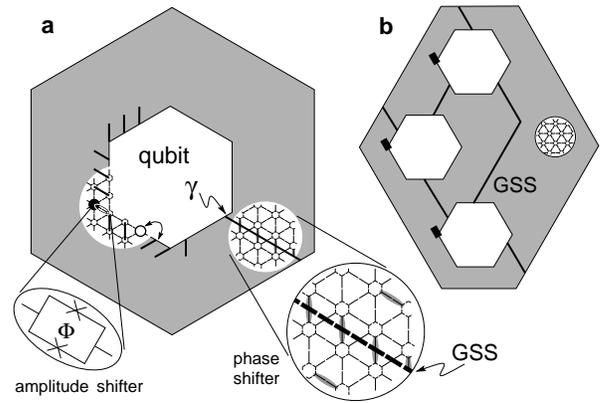}}
  \narrowtext\hspace{0.0cm}
  \caption{{\bf a} The topologically protected qubit involves the
  two degenerate groundstates of the quantum dimer liquid state
  realized within a triangular Josephson junction array. The
  qubit's phase is manipulated through the gated superconducting
  strip (GSS) attracting dimers along $\gamma$. The amplitude
  shifter mixing the groundstates is realized with the help of
  a tunable Josephson junction placed at the inner qubit boundary.
  Dimer-flips, like a row of falling dominos, take the hole around
  the inner boundary. {\bf b} Array of topologically protected
  qubits embedded in a Josephson junction array emulating a
  dimer system. The entanglement of two qubits proceeds through
  coupling over a gated superconducting strip.}
  \label{fig:qubit}
\end{figure}

We construct a topologically protected qubit from the two-level
system defined by the two ground states of a quantum
dimer system with cylindrical boundary conditions. We
make use of the above Josephson junction arrays and a
ring geometry; the two qubit states $|e\rangle$ and
$|o\rangle$ then are distinct through the parity of the
dimer count along the line $\gamma$ joining the inner
and outer boundaries of the array, see Fig.\ \ref{fig:qubit}.
In order to manipulate the qubit we have to implement the qubit
Hamiltonian
\begin{equation}
   H_{\rm qubit} = h_x \sigma_x + h_z \sigma_z,
   \label{H_qubit}
\end{equation}
where $\sigma_x$ and $\sigma_z$ are Pauli matrizes and $h_x$,
$h_z$ are the (manipulable) parameters producing the amplitude
($\alpha$) and phase ($\chi$) mixing in the qubit state $|\alpha,
\chi\rangle=[\,|e\rangle+\alpha\exp(i\chi) |o\rangle\,]/
\sqrt{1+\alpha^2}$. The implementation of $h_x$ requires a
controlled mixing of the protected dimer states, implying a
reduction of the ground state's topological protection. In the JJT
array this is achieved by breaking one dimer bond and creation of
a virtual particle-hole excitation where one Cooper pair retreats
to one hexagon (particle), leaving the partner hexagon empty
(hole). This virtual excitation costs the energy $\tilde{E}^{\rm
\scriptscriptstyle J}_l$ of that particular bond and takes the
system (virtually) out of the protected dimer space. While the
particle remains pinned to the weak junction, the hole is taken
around the inner boundary through appropriate dimer flips and is
subsequently recombined with the particle. This process results in
a mixing amplitude $h_x \sim E^{\rm \scriptscriptstyle J}_h
(E^{\rm\scriptscriptstyle J}_h/ \tilde{E}^{\rm \scriptscriptstyle
J}_l)^{M}$, where $M$ denotes the number of links on the inner
boundary (this estimate applies to the `optimal' dimer
configuration shown in Fig.\ \ref{fig:qubit}). Hence introducing
one switchable link junction near the inner boundary allows us to
tune the parameter $\tilde{E}^{\rm \scriptscriptstyle J}_l$ and
change the mixing amplitude $h_x$ by many orders of magnitude. The
variant for the JJK array involves a virtual excitation with two
Cooper pairs on one hexagon.
\vspace{-0.3cm}

The `phase shifter' can be implemented through a gated
superconducting strip, capacitively coupled to the array and
attracting dimers onto the reference line $\gamma$. Upon biasing
the strip, the energies of the two groundstates are shifted with
respect to one another and the qubit's phase $\chi$ is modified
accordingly. The implementation requires careful adiabatic
switching, with the amplitude $u$ (the energy acting on one dimer)
and the duration $\tau$ of the manipulation limited by the
constraints $u \sim t \ll E^{\rm\scriptscriptstyle J}_l$ and $\tau >
\hbar/\Delta$ in order to avoid excitations within the dimer
liquid. Furthermore, we have to break up the strip during idle
time, as the fully connected strip represents a global operator
(otherwise, electric fluctuations fed to the strip would decohere
the system). Thus the strip has to be constructed from isolated
superconducting islands which are to be connected only during the
time where the phase shifter becomes operational. Fast switches
accomplishing such a connection can be built from superconducting
Cooper pair transistors \cite{SSET}.

The qubit's manipulation involve strong modifications
of the system: the amplitude shifter requires leaving the protected
Hilbert space through virtual breaking of a dimer, while the phase
shifter requires introduction of a global operator. In turn,
these processes are strongly inhibited during idle time and give the
qubit its robustness: amplitude mixing is exponentially small in $L_x$
while phase drifting is exponentially small in $L_y$, cf.\ (\ref{phi_eo}).

Constructing a register of qubits is surprisingly simple: the same
basic Josephson junction array can accommodate $K$ qubits in a
geometry with $K$ holes, see Fig.\ \ref{fig:qubit}. Individual
qubit operations are carried out as described above. The
implementation of a non-trivial two-qubit operation again involves
a superconducting strip: biasing the strip connecting two qubits
lifts the energy of the states $|e\, o \rangle\,$ and $|o\,
e\rangle$ with respect to the states $|e\, e \rangle$ and $|o\,
o\rangle$ and combining this two-qubit `phase shifter' with a
suitable set of single qubit operations allows for the
construction of the controlled-NOT operation \cite{Ioffe}.

In a first step towards the implementation of topologically
protected qubits one may wish to test the proper functionality of
the quantum Josephson junction array. Such a test is provided by a
measurement of the magnetic susceptibility $\chi$ of the
ring-shaped structure: applying a magnetic field, the mixing
energy $h_x$ picks up a factor $\cos(2\pi \Phi/\Phi_0)$, where
$\Phi_0 = hc/e$ is the normal state flux quantum. This factor
originates from the Aharonov-Bohm phase picked up by the virtual
charged excitations encircling the hole. In the absence of
disorder the ring's energy changes with flux, $E_{\rm ring}(\Phi)
= h_x |\cos(2\pi \Phi/\Phi_0)|$, resulting in a suceptibility
$\chi \propto \partial_\Phi^2 E_{\rm ring}$ with period $\Phi_0/2$
and a sharp feature at $\Phi = \Phi_0/4$. The splitting $\Delta_d$ 
in the groundstate energies (due to disorder or finite-size effects)
smears this sharp feature over a region $\delta \Phi/\Phi_0 \sim
\Delta_d/h_x$; hence a measurement of $\chi(\Phi)$ allows to check
on the mixing amplitude $h_x$ and on the level splitting.

A particular challenge faced by qubit implementations based on
Josephson junctions operating in the charge limit is the presence
of stray charges \cite{Shnirman,Nakamura_99}; the robustness in
our device efficiently suppresses fluctuations in the gate
potential once they drop below $t$. The problem of stray charges
is avoided in designs using the dual phase variable
\cite{Mooij,Ioffe,Wal_00} and work in this direction is in
progress.

We thank V.\ Geshkenbein for discussions and acknowledge financial
support through the SCOPES program (swiss federal department of
foreign affairs and SNF), the NWO-Russia collaboration program,
the RFBR grant 01-02-17759, the program `Quantum Macrophysics' of
the Russian Academy of Science, and the Russian Ministry of
Science. Computations have been carried out on the Beowulfcluster
Asgard at ETHZ.

\end{multicols}

\end{document}

%% file: vv.eepic
\setlength{\unitlength}{0.00004in}
\begingroup\makeatletter\ifx\SetFigFont\undefined%
\gdef\SetFigFont#1#2#3#4#5{%
  \reset@font\fontsize{#1}{#2pt}%
  \fontfamily{#3}\fontseries{#4}\fontshape{#5}%
  \selectfont}%
\fi\endgroup%
{\renewcommand{\dashlinestretch}{30}
\begin{picture}(1692,1707)(0,-10)
\thicklines
\path(171,171)(171,1521)
\path(1521,1521)(1521,171)
\end{picture}
}

%% file: hh.eepic
\setlength{\unitlength}{0.00004in}
\begingroup\makeatletter\ifx\SetFigFont\undefined%
\gdef\SetFigFont#1#2#3#4#5{%
  \reset@font\fontsize{#1}{#2pt}%
  \fontfamily{#3}\fontseries{#4}\fontshape{#5}%
  \selectfont}%
\fi\endgroup%
{\renewcommand{\dashlinestretch}{30}
\begin{picture}(1692,1707)(0,-10)
\thicklines
\path(1521,1521)(171,1521)
\path(1521,171)(171,171)
\end{picture}
}

%% file: hh2.eepic
\setlength{\unitlength}{0.00004in}
\begingroup\makeatletter\ifx\SetFigFont\undefined%
\gdef\SetFigFont#1#2#3#4#5{%
  \reset@font\fontsize{#1}{#2pt}%
  \fontfamily{#3}\fontseries{#4}\fontshape{#5}%
  \selectfont}%
\fi\endgroup%
{\renewcommand{\dashlinestretch}{30}
\begin{picture}(3042,1707)(0,-10)
\thicklines
\path(1521,171)(171,171)
\path(2871,1521)(1521,1521)
\end{picture}
}

%% file: dd1.eepic
\setlength{\unitlength}{0.00004in}
\begingroup\makeatletter\ifx\SetFigFont\undefined%
\gdef\SetFigFont#1#2#3#4#5{%
  \reset@font\fontsize{#1}{#2pt}%
  \fontfamily{#3}\fontseries{#4}\fontshape{#5}%
  \selectfont}%
\fi\endgroup%
{\renewcommand{\dashlinestretch}{30}
\begin{picture}(3042,1707)(0,-10)
\thicklines
\path(171,171)(1521,1521)
\path(1521,171)(2871,1521)
\end{picture}
}

%% file: vv2.eepic
\setlength{\unitlength}{0.00004in}
\begingroup\makeatletter\ifx\SetFigFont\undefined%
\gdef\SetFigFont#1#2#3#4#5{%
  \reset@font\fontsize{#1}{#2pt}%
  \fontfamily{#3}\fontseries{#4}\fontshape{#5}%
  \selectfont}%
\fi\endgroup%
{\renewcommand{\dashlinestretch}{30}
\begin{picture}(1692,3057)(0,700)
\thicklines
\path(1521,2871)(1521,1521)
\path(171,171)(171,1521)
\end{picture}
}

%% file: dd2.eepic
\setlength{\unitlength}{0.00004in}
\begingroup\makeatletter\ifx\SetFigFont\undefined%
\gdef\SetFigFont#1#2#3#4#5{%
  \reset@font\fontsize{#1}{#2pt}%
  \fontfamily{#3}\fontseries{#4}\fontshape{#5}%
  \selectfont}%
\fi\endgroup%
{\renewcommand{\dashlinestretch}{30}
\begin{picture}(1692,3057)(0,700)
\thicklines
\path(171,1521)(1521,2871)
\path(171,171)(1521,1521)
\end{picture}
}

%% file: d.eepic
\setlength{\unitlength}{0.00004in}
\begingroup\makeatletter\ifx\SetFigFont\undefined%
\gdef\SetFigFont#1#2#3#4#5{%
  \reset@font\fontsize{#1}{#2pt}%
  \fontfamily{#3}\fontseries{#4}\fontshape{#5}%
  \selectfont}%
\fi\endgroup%
{\renewcommand{\dashlinestretch}{30}
\begin{picture}(1692,1707)(0,-10)
\thicklines
\path(1521,1521)(171,171)
\end{picture}
}

%% file: qc_dim_sh_corr.bbl
\vspace{-0.3cm}
\begin{thebibliography}{99}
\vspace{-1.3cm}

\bibitem{Cirac-Zoller} Cirac, J.I.\ \& Zoller, P.\
   Quantum computations with cold trapped ions.
   {\it Phys.\ Rev.\ Lett.\ }{\bf 74}, 4091-4094 (1995).

\bibitem{Monroe} Monroe, C., Meekhof, D., King, B., Itano, W.\
   \& Wineland, D.
   Demonstration of a fundamental quantum logic gate.
   {\it Phys.\ Rev.\ Lett.\ }{\bf 75}, 4714-4717 (1995).

\bibitem{Turchette} Turchette, Q., Hood, C., Lange, W., Mabushi,
   H.\ \& Kimble, H.J.
   Measurement of conditional phase shifts for quantum logics.
   {\it Phys.\ Rev.\ Lett.\ }{\bf 75}, 4710-4713 (1995).

\bibitem{Loss} Loss, D.\ \& DiVincenzo, D.P.\
   Quantum computation with quantum dots.
   {\it Phys.\ Rev.\ A} {\bf 57}, 120-126 (1998).

\bibitem{Shnirman} Shnirman, A., Sch\"on, G.\ \& Hermon, Z.\
   Quantum manipulations of small Josephson junctions.
   {\it Phys.\ Rev.\ Lett.\ }{\bf 79}, 2371-2374 (1997).

\bibitem{Averin} Averin, D.V.\
   Adiabatic quantum computation with Cooper pairs.
   {\it Solid State Commun.\ }{\bf 105}, 659-664 (1998).

\bibitem{Mooij} Mooij, J.E., Orlando, T.P, Levitov, L.S., Tian, L.,
   van der Wal, C.H. \& Lloyd, S.\
   Josephson persistent-current qubit.
   {\it Science} {\bf 235}, 1036-1039 (1999).

\bibitem{Ioffe} Ioffe, L., Geshkenbein, V.B., Feigel'man, M.V.,
   Fauch\`ere, A.L. \& Blatter, G.
   Environmentally decoupled $s$-wave--$d$-wave--$s$-wave Josephson
   junctions for quantum computing,
   {\it Nature} {\bf 398}, 678-681 (1999).

\bibitem{Nakamura_99} Nakamura, Y., Pashkin, Yu.A. \& Tsai, J.S.
   Coherent control of macroscopic quantum states in a
   single-Cooper-pair box.
   {\it Nature} {\bf 398}, 786-788 (1999).

\bibitem{Friedmann_00} Friedman, J.R., Patel, V., Chen,
   W., Tolpygo, S.K. \& Lukens, J.E.
   Quantum superposition of distinct macroscopic states.
   {\it Nature} {\bf 406}, 43-46 (2000).

\bibitem{Wal_00} van der Wal, C.H., ter Haar, A.C.J., Wilhelm,
   F.K., Schouten, R.N., Harmans, C.J.P.M., Orlando, T.P., Lloyd,
   S.\ \& Mooij, J.E.
   Quantum superposition of macroscopic persistent-current states.
   {\it Science} {\bf 290}, 773-777 (2000).

\bibitem{Kitaev} Kitaev, A.Yu.
   Fault-tolerant quantum computation by anyons.
   quant-ph/9707021.

\bibitem{Preskill} Preskill, J.
   Fault-tolerant quantum computation.
   {\it Introduction to quantum computation and information}
   (World Scientific, Singapore, 1998).

\bibitem{Kivelson_87} Kivelson, S.A., Rokhsar, D.S. \& Sethna,
   J.P.
   Topology of the resonating valence-bond state: solitons and
   high-$T_c$ superconductivity.
   {\it Phys.\ Rev.\ B} {\bf 35}, 8865-8868 (1987).

\bibitem{Rokhsar_88} Rokhsar, D.S. \& Kivelson, S.A.
   Superconductivity and the quantum hard-core dimer gas.
   {\it Phys.\ Rev.\ Lett.\ } {\bf 61}, 2376-2379 (1988).

\bibitem{Wen_91} Wen, X.G.
   Mean-field theory of spin-liquid states with finite
   energy gap and topological orders.
   {\it Phys.\ Rev.\ B} {\bf 44}, 2664-2672 (1991).

\bibitem{Moessner_00} R.\ Moessner, R. \& Sondhi, S.L.
   Resonating valence bond phase in the triangular lattice quantum
   dimer model.
   {\it Phys.\ Rev.\ Lett.} {\bf 86}, 1881-1884 (2001).

\bibitem{Misguich_99} Misguich, G.,  Lhuillier, C., Bernu, B.
   \& Waldtmann, C.
   Spin-liquid phase of the multiple-spin exchange Hamiltonian on the
   triangular lattice.
   {\it Phys.\ Rev.\ B} {\bf 60}, 1064-1074 (1999).

\bibitem{Sachdev_92} Sachdev, S.
   Kagome-acute- and triangular-lattice Heisenberg antiferromagnets.
   {\it Phys.\ Rev.\ B} {\bf 45}, 12377 (1992).

\bibitem{Fradkin} Fradkin, E.
   {\it Field Theories of Condensed Matter Systems}.
   Addison-Wesley, Redwood City, CA, 1991.

\bibitem{MSF} Moessner, R., Sondhi, S.L. \& Fradkin, E.
   Short-ranged RVB physics, quantum dimer models and Ising gauge
   theories.
   cond-mat/0103396.

\bibitem{Anderson} Anderson, P.W.
   The resonating valence bond state in La$_2$CuO$_4$ and
   superconductivity.
   {\it Science} {\bf 235}, 1196-1198 (1987).

\bibitem{Ioffe_89} Ioffe, L.B. \& Larkin, A.I.
   Superconductivity in the liquid-dimer valence-bond state.
   {\it Phys.\ Rev.\ B} {\bf 40}, 6941-6947 (1989).

\bibitem{FradkinShenker} Fradkin, E. \& Shenker, S.H.
   Phase diagrams of lattice gauge theories with Higgs fields.
   {\it Phys.\ Rev.\ D} {\bf 19}, 3682 (1979).

\bibitem{SSET} Grabert, H. \& Devoret, M.H.
   {\it Single charge tunneling: coulomb blockade phenomena in
   nanostructures}.
   Plenum Press, New York, 1992.

\end{thebibliography}
